\documentclass[aps,prd,twocolumn,nofootinbib,showpacs,superscriptaddress]{revtex4-1}

\usepackage{amssymb}
\usepackage{graphicx}
\usepackage{amsmath, amsthm}
\usepackage{epstopdf}
\usepackage{hyperref}

\newcommand{\be}{\begin{equation}}
\newcommand{\ee}{\end{equation}}

\begin{document}

\title{The charged McVittie spacetime} 

\author{Valerio Faraoni} 
\email[]{vfaraoni@ubishops.ca}

\author{Andres F. Zambrano Moreno}
\email[]{azambrano07@ubishops.ca}

\author{Angus Prain}
\email[]{aprain@ubishops.ca}
\affiliation{Physics Department and STAR Research Cluster, 
Bishop's University\\
Sherbrooke, Qu\'ebec, Canada J1M~1Z7
}

\begin{abstract} 

The two-parameter charged McVittie solution of the Einstein 
equations is revisited and its apparent horizons are discussed and 
located numerically (for the extremal case, analytically). 
According to the parameter values, this spacetime can be 
interpreted as a black hole, or a spacelike naked singularity, in a 
spatially 
homogeneous and isotropic universe.

\end{abstract}


\maketitle

\section{Introduction}

Analytical solutions of general relativity and of alternative theories of 
gravity which represent inhomogeneous cosmologies have been the subject of 
much recent interest. In the context of general relativity, spherically 
symmetric solutions of this kind are used to explore alternatives to dark 
energy in explaining the current acceleration of the cosmic expansion (see 
Ref.~\cite{voids} for a review), to study the effect of the cosmic dynamics 
on local systems \cite{CarreraGiulini}, to study the spatial variation of 
fundamental constants \cite{Barrow}, and to explore the thermodynamics of 
time-dependent horizons \cite{SaidaHaradaMaeda, myPRD, Majhi, 
myreview}. Solutions 
representing time-varying black holes are of interest in themselves and the 
first spacetime of this kind is the 1933 McVittie solution of the Einstein 
equations constructed to study the effect of the cosmic dynamics on a local 
system \cite{McVittie}. It describes a central inhomogeneity in a 
Friedmann-Lem\^aitre-Robertson-Walker (FLRW) ``background''.
 Over the years, the McVittie solution has been the subject of many studies
\cite{oldMcVittie}  
but it is not yet completely understood, as testified by the proliferation 
of recent works \cite{CarreraGiulini10, 
Klebanetal, LakeAbdelqader, Anderson, Roshinaetal, AndresRoshina, 
SilvaFontaniniGuariento}. 
Most recently, it has been shown that the 
McVittie metric cannot be generated as a scalar field solution unless 
non-canonical scalars, introduced in the literature as forms of exotic dark 
energy, are employed --- this is the case of the cuscuton theory, a 
special case of Horava-Lifschitz theory, of which 
the McVittie metric is a solution \cite{cuscuton}.

A charged version of the original McVittie metric was found by Gao and Zhang 
\cite{GaoZhang}, who also generalized it to higher dimensions 
\cite{McVittieHigherD}, and was discussed very briefly in 
\cite{McClureDyerCQG, McClureThesis}. Here we want to refine the rather 
cursory analysis of this metric presented in these references and, in 
particular, locate its apparent horizons (when present) and study their 
dynamics. 

Conformal diagrams of the McVittie spacetime for various backgrounds were obtained 
in \cite{LakeAbdelqader, SilvaFontaniniGuariento} and we expect them to be 
qualitatively similar in the charged case; however we do not get 
into such level 
of detail here. As an application of the charged McVittie spacetime, we mention 
that it was used\footnote{The analysis of Ref.~\cite{bhareas} is flawed by a 
typographical error present in the metric of \cite{GaoZhang} but, as we shall see, 
the qualitative behaviour of the apparent horizons for $|Q| \leq m$ does not change 
and the argument of Ref.~\cite{bhareas} still stands.}
to disprove the 
universality of certain quantization laws for quantities 
constructed with the areas of black hole apparent horizons and inspired 
by string theories \cite{bhareas}.

\section{The charged McVittie spacetime and its apparent horizons}

For simplicity we restrict ourselves to a spatially flat FLRW ``background'' 
(we use quotation marks because, due to the non-linearity of the Einstein 
equations, it is in general impossible to split a metric into a background and a deviation from it in a covariant way). The 
spherically symmetric, non-stationary, charged McVittie line element and the 
only nonvanishing component of the Maxwell tensor are\footnote{Our 
notations 
differ slightly from those of Ref.~\cite{GaoZhang}. Beware of a typographical error in
the numerator of $g_{00}$ in \cite{GaoZhang}, which was corrected in 
Refs.~\cite{McVittieHigherD, McClureDyerCQG, McClureThesis}.}
 \begin{eqnarray}
ds^2&=&-\frac{ \left[  1-\frac{ \left( m^2-Q^2 \right)}{4a^2r^2} \right]^2 }{
\left[ \left( 1+\frac{m}{2ar}\right)^2-\frac{Q^2}{4a^2r^2} \right]^2} \, dt^2 
\nonumber\\
&&\nonumber\\
&\, & +a^2(t)\left[ \left( 1+\frac{m}{2ar} \right)^2 
-\frac{Q^2}{4a^2r^2}\right]^2 \left( dr^2 
+r^2 d\Omega_{(2)}^2 \right) \,,\nonumber\\
&& \label{QMcVlineelement} \\
F^{01} &=& \frac{Q}{ a^2r^2\left[ 1-\frac{  \left( m^2-Q^2\right)}{4a^2r^2} \right]
\left[ \left( 1+\frac{m}{2ar}\right)^2 -\frac{Q^2}{4a^2r^2}\right]^2} \,,\nonumber\\
&&
\end{eqnarray}
where $r$ is the isotropic radius, $d\Omega_{(2)}^2=d\theta^2 +\sin^2\theta 
d\varphi^2$ is the line element on the unit 2-sphere,
 the constants $m>0$ and $Q$ are a mass
and an electric charge parameter, respectively, and 
$a(t) $ is the scale factor of the ``background'' FLRW universe.
If $a \equiv 1$, the line element (\ref{QMcVlineelement}) 
reduces to the Reissner-Nordstr\"om one 
in isotropic coordinates while, for large values of $r$, it reduces to 
the spatially flat FLRW metric. If $m=Q=0$, the metric becomes an exact 
spatially flat FLRW one. 

The inspection of eq.~(\ref{QMcVlineelement}) provides immediately 
the areal radius  
\begin{eqnarray} 
R(t,r) &=& a(t)r \left[ \left( 
1+\frac{m}{2a(t)r}\right)^2 -\frac{Q^2}{4a^2(t) r^2}\right] \nonumber\\
&&\nonumber\\
&=& m+a(t)r+\frac{m^2-Q^2}{4a(t)r} 
\end{eqnarray} 
and it is $R(t,r)\geq m $ for $|Q| \leq 
m$. Assuming the range of parameters $|Q| \leq m$, the function $R(t,r)$ 
(and consequently also the area $4\pi R^2$ of 2-spheres of symmetry)  
decreases from $+\infty$ in the range $ 0< ar< \sqrt{m^2-Q^2}/2$, has an 
absolute minimum $R_{min} =  
m+\sqrt{m^2-Q^2}$ at $ar=\sqrt{m^2-Q^2}/2$, 
and increases again to plus infinity for $ ar> 
\sqrt{m^2-Q^2}/2$, which shows clearly that the isotropic radius 
corresponds to a double covering of the spacetime region $ R> 
m+\sqrt{m^2-Q^2}\geq m $. (This fact is well known for the 
Schwarzschild 
spacetime \cite{Buchdhal}, which is contained in the metric 
(\ref{QMcVlineelement}) as the special case $a \equiv 1, Q=0$.)

In the case $|Q|\geq m$, the areal radius $R$ is instead a 
monotonically 
increasing function of $r$ and the physically meaningful region 
$R\geq 0$ corresponds to $r\geq \frac{|Q|-m}{2a(t)}\geq 0$.

The relation between areal radius $R$ and isotropic radius $r$ can be 
inverted, which will be useful in the following to study the horizons of 
the charged McVittie spacetime. This inversion gives the quadratic equation
\be
r^2-\frac{\left(R-m \right)}{a}\, r + \left( 
\frac{m^2-Q^2}{4a^2}\right)=0 \,,
\ee
with the positive root  satisfying the relation
\be \label{f(R)}
2ar = R-m +\sqrt{R^2+Q^2-2mR} \equiv f(R) \,.
\ee
The Ricci scalar is found to be
\begin{align} 
{R^c}_c &=  \frac{6}{ \left[ 1-\frac{\left(m^2-Q^2 \right)}{4a^2r^2} \right] }
\left\{ \frac{\ddot{a}}{a} \left[ \left( 1+\frac{m}{2 ar} \right)^2 - \frac{Q^2}{4a^2r^2}\right]
\right.\nonumber\\
&\,  \left. + H^2 \left[ 1-\frac{m}{ar} + \frac{3\left(Q^2 -m^2 
\right)}{4a^2r^2} \right] \right\} \, .
\label{Ricciscalar}
\end{align}
The non-zero components of the Einstein tensor are\footnote 
{The expression of the Einstein tensor in 
Refs.~\cite{McClureDyerCQG} and \cite{McClureThesis} differs from 
ours in that, there, a scale factor term is missing in the 
combination $[ m^2-Q^2+4mra+4r^2a^2]^4$ in the denominators.} 
\begin{align} 
{G^t}_t &= -\frac{256Q^2r^4a^4}{ [ m^2-Q^2+4mra+4r^2a^2]^4 
}+\frac{3\dot{a}^2}{a^2} \,,
\\{G^r}_r &= \frac{256Q^2r^4a^4}{ [ m^2-Q^2+4mra+4r^2a^2]^4 } 
\nonumber\\
&\,  
+\frac{\dot{a}^2(-5m^2+5Q^2-8mra+4r^2a^2)}{a^2(-m^2+Q^2+4r^2a^2)} 
\nonumber\\
&\,  +\frac{2\ddot{a}( m^2-Q^2+4mra+4r^2a^2)}{a(-m^2+Q^2+4r^2a^2)} 
\,,\\
{G^\theta}_\theta =  {G^\varphi}_\varphi &= -\frac {256Q^2r^4a^4}{ 
[ m^2-Q^2+4mra+4r^2a^2]^4 }\nonumber\\
&\,  +\frac{\dot{a}^2(-5m^2+5Q^2-8mra+4r^2a^2)}{a^2(-m^2+Q^2+4r^2a^2)}\nonumber\\
&\,  +\frac{2\ddot{a}( m^2-Q^2+4mra+4r^2a^2)}{a(-m^2+Q^2+4r^2a^2)} 
\,.
\label{EinTensor}
\end{align}
$H\equiv \dot{a}/a$ is the Hubble parameter of the FLRW 
``background'' and an overdot denotes differentiation with respect 
to the comoving time $t$. The vanishing of $G^{tr}$ implies that 
there is no accretion (which, to respect the symmetry,  could only 
be radial) of cosmic fluid onto the central object. 
 
For $m=Q=0$ eq.~(\ref{Ricciscalar}) reduces to the well known 
Ricci scalar of the spatially flat FLRW universe ${R^c}_c=6\left( 
\dot{H}+2H^2 \right)$. For $a\equiv 1$ it reduces to zero, which 
is the Ricci scalar of the Reissner-Nordstr\"om spacetime (the only 
source of matter is then the Maxwell field with traceless  
energy-momentum tensor). If $|Q|\leq m$ there is a singularity 
where $1-\frac{ \left(m^2-Q^2 \right)}{4a^2r^2}=0$, or 
\be 
ar= \frac{ \sqrt{m^2-Q^2} }{2} \,,
\ee
corresponding to the areal radius
\be\label{singularity}
R=m+\sqrt{m^2-Q^2} \,
\ee
which is precisely the location of the outer event horizon of Reissner-Nordstr\"om spacetime (the $a\equiv 1$ limit). 

For the extremal case $|Q|=m$ the metric becomes 
\be
ds^2=-\frac{dt^2}{\left(1+\frac{m}{ar}\right)^2}+a^2\left(1+\frac{m}{ar}\right)^2 dr^2+d\Omega_{(2)}^2
\ee
and the singularity occurs at $r=0$ which corresponds to the areal 
radius $R=m$, again coinciding with the location of the outer 
event horizon of the $a\equiv 1$ Reissner-Nordstr\"om limit, in 
this case the single event horizon of the extremal 
Reissner-Nordstr\"om black hole. 

This new spherical singularity is not present if $|Q|>m$. In this case, however, the scalar invariant of the 
Maxwell tensor
\be
F_{ab}F^{ab}= -\frac{ Q^2}{
a^2 r^4 \left[ \left( 1+\frac{m}{2ar} \right)^2 -\frac{Q^2}{4a^2r^2} \right]^2}
\ee
diverges at
\be
ar= \frac{ |Q|-m}{2} \,,
\ee
corresponding to $R=0$ due to the divergence of the radial electric 
field, which is the only non-zero component $F^{01}$ of the 
Maxwell tensor. Of course, the Big Bang singularity is always present at $a=0$. The singularity 
(\ref{singularity}) divides the spacetime into two disconnected 
manifolds, as in the McVittie case 
\cite{Nolan, oldMcVittie}.

The spacetime singularity (\ref{singularity}) (for $|Q| \leq m$) is 
spacelike. In fact, this  
singularity corresponds to $\psi=0$, where 
\be 
\psi(t,r) \equiv a(t)r -\frac{ \sqrt{m^2-Q^2}}{2} 
\,; 
\ee 
we have 
\begin{eqnarray} 
\nabla_c\psi \nabla^c \psi &=& -\dot{a}^2r^2 \frac{ \left[ \left( 
1+\frac{m}{2ar}\right)^2 -\frac{Q^2}{4a^2r^2}\right]^2}{ \left[ 1- \frac{ \left( m^2-Q^2 
\right)}{4a^2r^2}\right]^2 } \nonumber\\
&&\nonumber\\
&\, & +\frac{1}{\left[ \left( 1+\frac{m}{2ar}\right)^2 
-\frac{Q^2}{4a^2r^2}\right]^2} \,. 
\end{eqnarray} 
In the limit in which $2ar \rightarrow \sqrt{m^2-Q^2}$ from above, 
this expression tends to $-\infty$, therefore the norm of the normal to the surfaces 
$\psi=$const. in this limit is negative,  hence this surface and 
its limit are spacelike.

The black hole nature of the charged McVittie spacetime is assessed by examining its 
horizons. Since the metric is dynamical, the relevant concept of horizon 
is not the null and teleological event horizon, which requires the knowledge 
of the entire spacetime manifold to future null infinity. Apparent 
and 
trapping horizons, instead, are more appropriate and useful concepts 
to describe dynamical situations 
\cite{Alexreview, Ivanreview, AshtekarKrishnan, myreview}, 
and we will study the 
apparent horizons, which are located by the equation 
$\nabla^{c}R\nabla_{c}R=0$ ({\em e.g.}, \cite{NielsenVisser, 
AbreuVisser, myreview}), where $R$ is the areal radius. After 
straightforward manipulations, this equation becomes
\begin{eqnarray}
&& \left\{ \dot{a}^2r^2 \left[ \left( 
1+\frac{m}{2ar}\right)^2-\frac{Q^2}{4a^2r^2}\right]^4  \right. \nonumber\\
&&\nonumber\\
& \, & \left.  -\left[ 
1-\frac{\left( m^2 -Q^2 \right)}{4a^2r^2} \right]^2 \right\} \nonumber\\
&&\nonumber\\
&&\cdot \left[ \left( 1+\frac{m}{2ar}\right)^2 -\frac{Q^2}{4a^2r^2} \right]^{-2}=0 \,.
\label{mss}
\end{eqnarray}
Excluding the singularity~(\ref{singularity}), the roots of (\ref{mss}) 
are the roots of the equation
\be
\dot{a}^2r^2 \left[ \left( 
1+\frac{m}{2ar}\right)^2-\frac{Q^2}{4a^2r^2}\right]^4 =\left[ 
1-\frac{\left( m^2-Q^2 \right)}{4a^2r^2}  \right]^2 \,.
\ee
It is useful to express the location of the apparent horizons in 
terms of the physical (areal) radius, obtaining 
\be \label{AHlocation}  
4f^2R^4H^2 -\left( f^2-m^2 + Q^2 \right)^2=0 \,, 
\ee
where $f(R)$ is given by eq.~(\ref{f(R)}). Equation \eqref{AHlocation} can be manipulated without further approximation to the quartic polynomial in $R$
\be
H^2R^4-R^2+2mR-Q^2=0. \label{E:poly}
\ee
Note that this polynomial is general and applies independently of the relation of $Q$ to $m$. 
For large values of $R$ this equation reduces to $R \approx 1/H$, the 
value of the radius of the apparent horizon of the spatially flat FLRW ``background'', which 
makes us expect a cosmological horizon to be present. 
In any regime in which $H\rightarrow0$ (for example in power law 
backgrounds with $a(t)=a_0\, t^p$ and hence $H=p/t\rightarrow 0$ as 
$t\rightarrow\infty$), eq.~(\ref{AHlocation}) reduces to its 
asymptotic form  
\be
R^2-2mR+Q^2 \simeq 0 
\ee
so that, in this limit
\be
R=m\pm\sqrt{m^2-Q^2}
\ee
implying that at late times in any region of a background for which 
$H\rightarrow 0$ (for example late times in a power law background) a black hole apparent horizon asymptotes to the singularity 
(\ref{singularity}) from above. Note that the smaller root always 
lies inside the spherical singularity and that these two roots 
exactly coincide with the locations of the two event horizons of 
Reissner-Nordstr\"om. Hence we see that the limit $H\rightarrow 0$ 
reproduces the horizon structure of the static 
Reissner-Nordstr\"om  
spacetime.\footnote{Albeit with the addition of the spherical 
singularity not present in Reissner-Nordstr\"om.  This spherical 
singularity is present even when $|Q|=m$ and independently of the 
time dependence $a(t)$, as long as it is not exactly constant.  The 
authors are currently investigating further this curious feature  
of the charged and uncharged McVitte spacetime.}

\subsection{Dust-dominated background}

\begin{figure} \begin{center} 
\scalebox{0.45}{\rotatebox{0}{\includegraphics{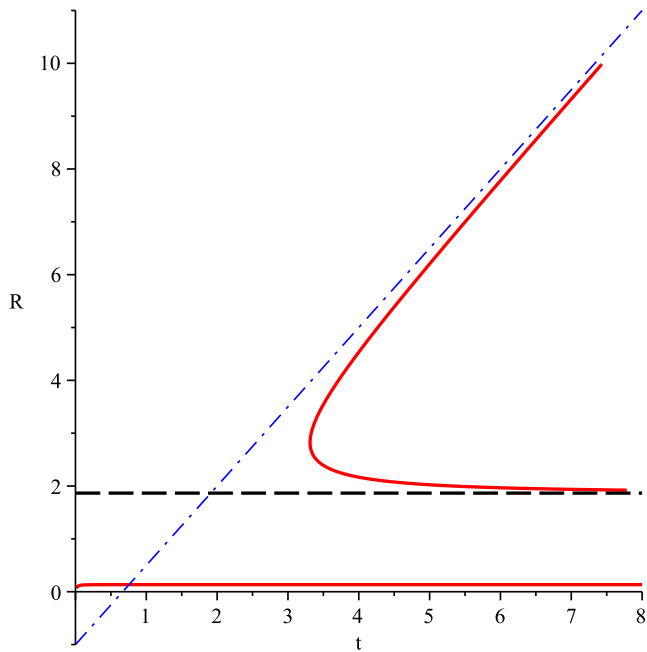}}} 
\end{center} 
\caption{The physical radii $R$  
of the apparent horizons of the charged McVittie 
spacetime versus time $t$ for $m=1$ and $Q=1/2$ and dust-dominated  
power law background $a(t)=t^{2/3}$. The blue dash-dotted line is 
the location of the corrected `Hubble radius' $R=1/H-m$ which is an asymptote 
for the largest apparent horizon at late times while the black dashed line is the location of the spherical singularity \eqref{singularity}. }
\label{AHradii} 
\end{figure}

Let us restrict now, for the sake of example, to a dust-dominated background with $a(t)=a_0 
t^{2/3}$ and $H(t)=\frac{2}{3t}$. Consider the case $|Q|<m$.  Then eq.~\eqref{E:poly} can be solved explicitly at different times $t$ giving the location of the apparent horizons.  These are plotted versus comoving time in 
Fig.~\ref{AHradii} for a choice of parameters. The  
phenomenology of these apparent horizons is similar to that found for the 
uncharged McVittie solution \cite{Nolan, Klebanetal, LakeAbdelqader, 
AndresRoshina}. 

Initially (near the Big Bang $t=0$), there are no apparent horizons in the connected outer region beyond the spherical singularity; at a critical 
time a cosmological and a black hole apparent horizon appear. The 
black hole apparent horizon shrinks asymptoting to the singularity, 
while the  cosmological apparent horizon expands forever. This 
phenomenology is 
interpreted in \cite{AndresRoshina} for $Q=0$ by noting that at early times the size 
of the black hole horizon exceeds that of the cosmological horizon and such 
a large black hole cannot be accomodated in a small universe. (This situation 
is similar to that of the Schwarzschild-de Sitter black hole, which is a 
special case of the McVittie metric \cite{AndresRoshina}.)  Later on, the 
cosmological horizon becomes larger than the black hole apparent horizon, 
which now fits below the cosmological horizon \cite{AndresRoshina}.

 Note 
that the innermost apparent horizon, which asymptotes to the 
location of the inner unstable Cauchy horizon of 
Reissner-Nordstr\"om geometry, is located in the inner disconnected 
region, separated from the exterior geometry by the singularity 
\eqref{singularity}, converging to $R=0$ at the Big Bang.

\subsection{Universe undergoing finite expansion}

\begin{figure}
\centering
\includegraphics[scale=0.45]{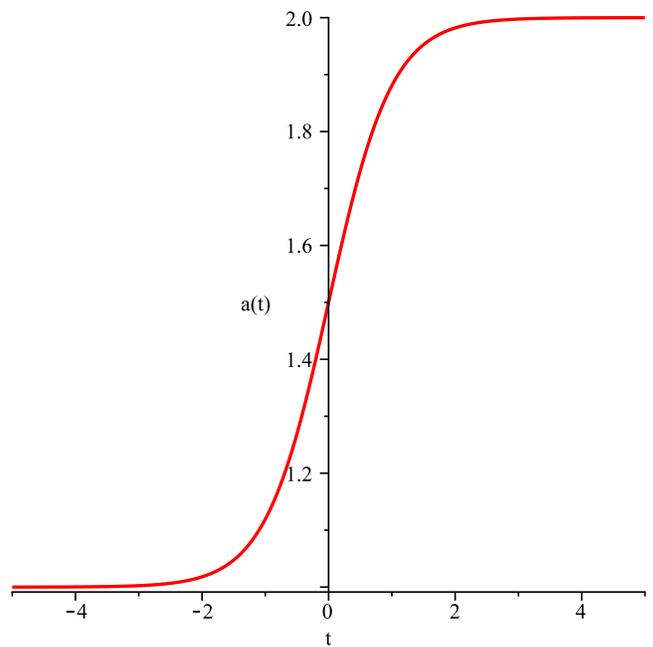}
\caption{ The finite expansion scale factor \eqref{E:finite} for 
$a_\text{i}=1$, $a_\text{f}=2$ and $t_0=1$.\label{F:finite}}
\end{figure}

As a toy model to further investigate the interesting apparent 
horizon structures, let us consider a universe undergoing a finite 
expansion from an initially quasi-static regime to a final 
quasi-static regime with scale factor
\be
a(t)=\frac{a_\text{f}+a_\text{i}}{2}+\frac{a_\text{f}-a_\text{i}}{2}\tanh\,\left(\frac{t}{t_0}\right). \label{E:finite}
\ee
Shown in Fig.~\ref{F:finite} is the scale factor \eqref{E:finite} 
for the  choice $a_\text{i}=1$, $a_\text{f}=2$, and $t_0=1$ 
representing a universe which expands by a factor of 2 from $t=-\infty$ to $t=+\infty$. We consider only the case $|Q|<m$ for the rest of this subsection and also fix the values $m=1$ and $Q=1/2$ for illustration. 

\begin{figure}
\centering
\includegraphics[scale=0.45]{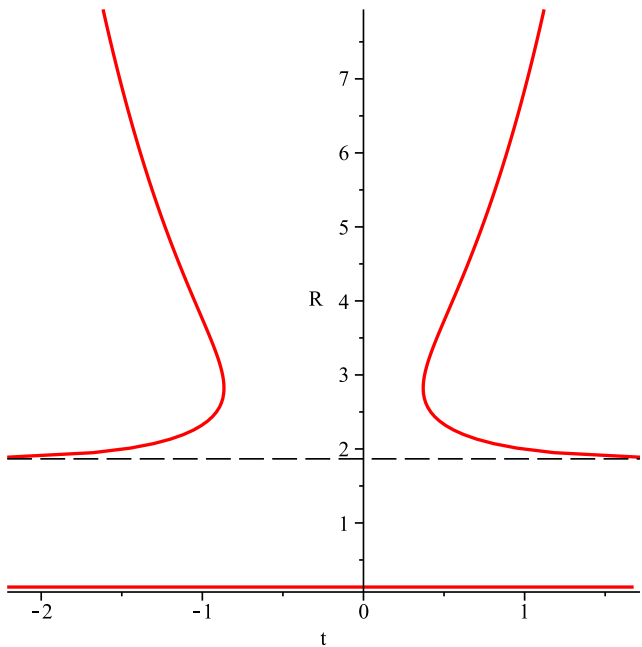}
\caption{The locations of the apparent horizons over time for the  
``fast'' finite expansion background with scale factor 
\eqref{E:finite} with $t_0=1$ given by the red solid curves. The 
black dashed line is the location of the spherical singularity.  
\label{F:fast_finite}}
\end{figure}

The horizon phenomenology in this case depends on the time scale of 
expansion $t_0$ with horizon mergers and horizon joinings possible. 
In Fig.~\ref{F:fast_finite} we plot the roots of the polynomial 
\eqref{E:poly} over time for a `fast' expansion ($t_0=1$) where the 
cosmological horizon shrinks from $R=+\infty$ to eventually merge 
with the expanding black hole horizon leaving a naked singularity 
for a period followed by the birth again of two horizons which 
separate, the cosmological one out to $R=+\infty$ and the black 
hole horizon asymptoting to the spherical singularity at 
$R=m+\sqrt{m^2-Q^2}$. 

\begin{figure}
\centering
\includegraphics[scale=0.45]{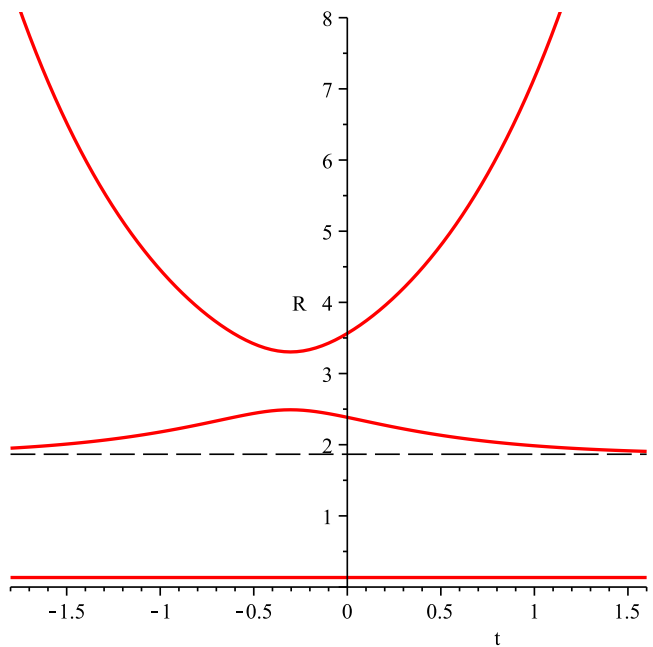}
\caption{The locations of the apparent horizons over time for the 
``slow'' finite expansion background with scale factor 
\eqref{E:finite} and $t_0=0.57$, given by the red solid curves. The 
black dashed line is the location of the spherical 
singularity.\label{F:slow_finite}}
\end{figure}

Note that the horizon structure is not symmetric about $t=0$ since 
it is the asymmetric function $H(t)=\dot{a}/a$ which enters the 
quartic polynomial defining the horizon locations. In 
Fig.~\ref{F:slow_finite} we plot the location of the apparent 
horizons over time for a ``slow'' expansion ($t_0=0.57$) where the 
cosmological horizon does not shrink enough to subsume the black 
hole horizon. In this slow case, the spherical singularity is 
always hidden behind an apparent horizon for asymptotic observers 
far from the inhomogeneity. 

\begin{figure}
\centering
\includegraphics[scale=0.45]{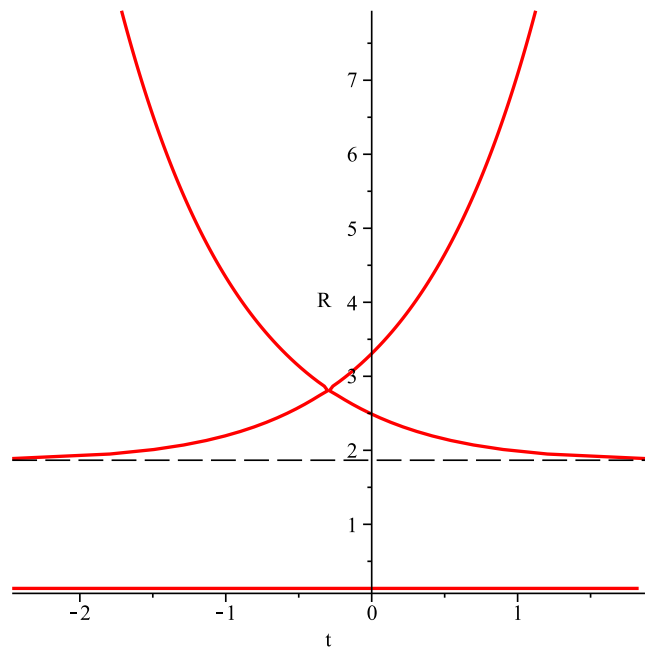}
\caption{The location of the apparent horizons in the limiting case 
in which the time scale of the finite expansion is critical, 
resulting in the momentary merger of the two outer horizons, 
appearing to cross. The black dashed line is the location of the 
spherical singularity.\label{F:curious_cross}}
\end{figure}

Interestingly, the curious limiting case in which the cosmological 
and  black hole horizons meet at a point, appearing to intersect, 
can be solved for exactly, occurring when the Hubble 
parameter attains a critical value $H_\text{crit}$ at its maximum 
leaving a single apparent horizon at $R_\text{crit}$ in the 
exterior region, where
\begin{align}
R_\text{crit}&=\frac{3}{2}m+\frac{1}{2}\sqrt{9m^2-8Q^2} \,, \\ 
H_\text{crit}&=\frac{1}{R_\text{crit}}\sqrt{\frac{1}{3}
-\frac{Q^2}{3R_\text{crit}}} 
\,.
\end{align}
Such a scenario is shown in Fig.~\ref{F:curious_cross}, where we 
see the ``crossing'' of the horizons at a single time.

\subsection{The extremal case $|Q|=m$}

\begin{figure}
\centering
\includegraphics[scale=0.45]{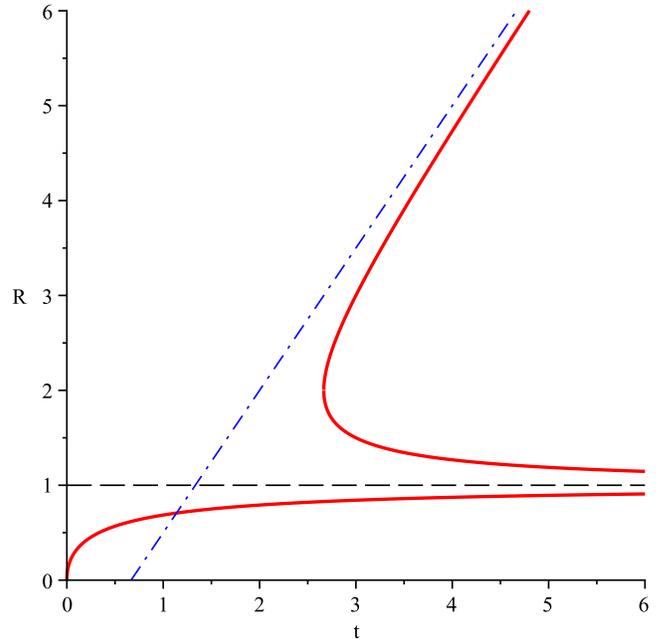}
\caption{The location of the apparent horizons over time for the 
dust-dominated extremal case $|Q|=m$ and $a(t)\propto t^{2/3}$. The 
blue dash-dotted curve is the location of the corrected Hubble radius  
$R=1/H-m$, while the dashed black line is the location of the single 
RN event horizon $R=m$ to which the two inner apparent horizons 
asymptote at late times. $R=m$ is also the location of the 
spherical singularity \eqref{singularity}.  \label{F:extremal}}
\end{figure}

In the extremal case with $|Q|=m$ the quartic relation 
\eqref{E:poly} is exactly solvable, with solutions
\be
R_\text{extremal}:=\frac{ 1}{2H} \pm\frac{\sqrt{1-4mH}}{2H}, \, \frac{ -1}{2H} \pm\frac{\sqrt{1+4mH}}{2H}. \label{explicit}
\ee
The extremal case $|Q|=m$ is interesting in the sense that 
an explicit analytical expression like (\ref{explicit}) for the 
apparent horizon radii is rare to find in investigations of 
time-evolving black holes \cite{myreview}. One of these roots is 
always negative and hence is discarded
\be
R_\text{negative}=-\frac{1}{2H}-\frac{\sqrt{1+4mH}}{2H}.
\ee
The smallest positive root 
given by 
\be
R_\text{inner}:=-\frac{ 1}{2H}+ \frac{\sqrt{1+4mH}}{2H}
\ee
is always present while the other two
\be
R_{\pm}:=\frac{ 1}{2H}\pm \frac{\sqrt{1-4mH}}{2H}
\ee
can merge (become complex) or appear simultaneously (become real) 
depending on the evolution of $H$.  Interestingly, for the 
dust-dominated background $H\rightarrow 0$ as $t\rightarrow 
+\infty$ and we see that  $R_\text{inner}$ and $R_{-}$ in fact 
converge to the same radius $R=m$. This common limit radius is 
precisely where the single event horizon of extremal 
Reissner-Nordstr\"om sits, nicely showing that we recover the 
phenomenology of the static charged spacetime in the appropriate 
limit. This horizon behaviour is shown in Fig.~\ref{F:extremal}.

\subsection{The supercritical case $|Q|>m$}

\begin{figure} \begin{center} 
\scalebox{0.45}{\rotatebox{0}{\includegraphics{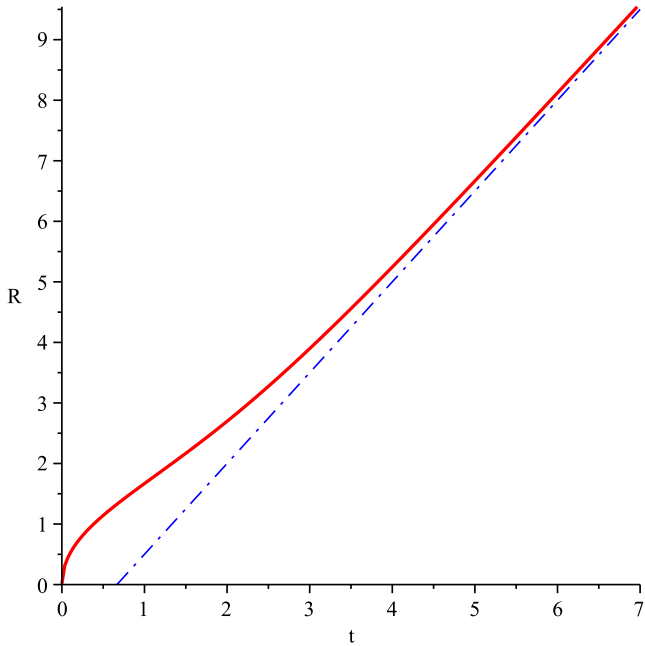}}} 
\end{center} 
\caption{The areal radius 
of the apparent horizon of the charged McVittie spacetime versus 
time for $|Q|>m$ in solid red. The dashed blue line is the corrected Hubble 
radius $R=1/H-m$ to which the single apparent horizon asymptotes from 
above at late times motivating its characterisation as a 
cosmological apparent horizon. There is a naked singularity at 
$R=0$.}
\label{AHradii2} 
\end{figure}

In the case $|Q|>m$ we expect a naked singularity, based on the limit of constant $a$ which 
reproduces the Reissner-Nordstr\"om spacetime. The cosmological horizon is still present, as 
expected from the previous considerations. In fact, numerical plots for this situation  
provide a single (cosmological) apparent horizon, as shown in 
Fig.~\ref{AHradii2}.

\section{Discussion}

In the parameter range $|Q| \leq m$ explored we find only a cosmological apparent 
horizon and one black hole apparent horizon, while in the 
Reissner-Nordstr\"om black hole (to which the metric reduces if $a \equiv 
1$) it is well known that also an inner black hole (apparent) horizon is present. This 
fact begs the question of why, in the charged McVittie spacetime which generalizes the 
Reissner-Nordstr\"om metric, we do not 
see an inner  black hole horizon 
in addition to the cosmological horizon. The inner black hole horizon 
of the Reissner-Nordstr\"om black hole is unstable with respect to linear 
perturbations and the effect of the cosmological ``background'' on 
the central inhomogeneity is akin to a non-linear (or exact) ``large 
perturbation'', therefore, the absence of an inner black hole 
horizon is not too surprising in this regard.\footnote{However, 
this ``perturbation'' is not ``small'', in the sense 
that it corresponds to an infinite amount of  mass-energy added to 
the Reissner-Nordstr\"om spacetime, even in the 
case of arbitrarily small density $\rho$ of the FLRW 
``background''. Moreover,  the density $\rho$  
diverges at the spherical singularity.} We expect such 
an horizon to be absent in any exact solution of the Einstein 
equations describing a Reissner-Nordstr\"om-like black hole 
interacting with a non-trivial environment.

\begin{acknowledgments} 
{\footnotesize This work is supported by Bishop's University 
and by the Natural Sciences and Engineering Research Council of Canada.}
\end{acknowledgments}


\end{document}